\documentclass[letter]{aa}
\usepackage{amssymb,amsmath}
\usepackage{graphicx}
\usepackage{xcolor,natbib}
\usepackage{multirow}
\usepackage[T1]{fontenc}
\usepackage{ae,aecompl}
\usepackage{newtxtext, newtxmath}
\usepackage{float}
\usepackage{subfigure}
\usepackage{longtable}
\usepackage{enumitem}
\usepackage{longtable,listings}
\usepackage[flushleft]{threeparttable}
\usepackage{parcolumns}
\usepackage{hyperref}

\def\oc3{[O~{\sc iii}]$_c$}
\def\ob3{[O~{\sc iii}]$_b$}
\def\obj{AT2020wey}

\begin{document}

\title{Similar ratios of rise timescale to decline timescale of optical light curves in common tidal disruption events}

\titlerunning{A classification parameter for optical TDEs}
\authorrunning{Zhang}

\author{XueGuang Zhang}

\institute{Guangxi Key Laboratory for Relativistic Astrophysics, School of Physical Science 
and Technology, GuangXi University, No. 100, Daxue Road, Nanning, 530004, 
P. R. China \ \ \ \email{xgzhang@gxu.edu.cn}}

\abstract{Totally similar physical process in tidal disruption events (TDEs) basically indicates that there should be potential 
parameter to distinguish variability properties of TDEs from the other transient events having different physical processes. Here, 
we try to report such a parameter, the timescale ratio $R_{2/1,rd}$ of rise timescale $t_{1/2,r}$ (from half-max to maximum) to 
decline timescale $t_{1/2,d}$ (from maximum to half-max), especially based on the 34 optical TDEs with reported $t_{1/2,r}$ and 
$t_{1/2,d}$. Among the 34 optical TDEs, AT2020wey is an outlier with $R_{2/1,rd}\sim2.7$ which is 4.5 times larger than the mean 
value 0.6 of the other optical TDEs. However, after considering similar but more flexible model functions, the re-determined 
$R_{2/1,rd}$ is $\sim$0.9 in AT2020wey, totally similar as the values of the other optical TDEs. Therefore, the parameter 
$R_{1/2,rd}\sim0.6$ could be a potential classification parameter for optical TDEs. Furthermore, $R_{1/2,rd}$ have been checked 
in the unique optical transients of AT2019avd, PS1-10adi, SDSS J0946+3512 and J2334+1457. We can find that the second flare with 
$R_{1/2,rd}\sim11$ in AT2019avd should be very different from the other optical TDEs, but PS1-10adi, SDSS J0946+3512, J2334+1457 
and the first flare in AT2019avd should be similar as the other optical TDEs. In the near future, properties of $R_{1/2,rd}$ 
through large sample of optical transients could provide further clues to support whether $R_{1/2,rd}$ could be a better 
classification parameter to distinguish TDEs and the other transient events.}


\maketitle

\section{Introduction}

	Tidal disruption events (TDEs) have been studied for more than four decades since the pioneer work in 1980s by \citet{re88, 
ek89}, considering a star tidally disrupted by a central supermassive black hole (SMBH). There are so far more and more brilliant 
works on both theoretical simulations and observational reports on TDEs which have been commonly accepted as candles for central 
SMBHs and corresponding BH accreting systems in galaxies. More recent theoretical simulations on TDEs can be found in \citet{gr13, 
gm14, lf15, sk18, cn19, bl20, cn20, lo21, tk22, ks23, zh23a, rm24, pk25, yl25, ym25}, etc., leading to theoretical model expected 
variability patterns in different wavelength bands with considerations of different physical environments. Then, theoretical 
simulations provide evident clues for detecting optical TDEs through model expected variability patterns in optical light curves, 
leading to so far around 400 optical TDEs reported in the literature, such as the reported samples of TDEs candidates in \citet{wy18, 
sg21, vg21, hv23, yr23, zh25a}. Recent review on optical TDEs candidates can be found in \citet{ref2}.

	Besides optical TDEs light curves expected and described by theoretical TDE models, mathematical methods have also been 
applied to describe the observed optical light curves, such as the more recent detailed descriptions in \citet{hv23, yr23, yr25}. 
Applications of mathematical methods can lead to some interesting parameters of the profile properties of the optical light curves, 
especially the determined timescales of $t_{1/2,r}$ (the rise timescale from half-max to maximum) and $t_{1/2,d}$ (the decline 
timescale from maximum to half-max). In \citet{yr23, yr25}, the parameters of $t_{1/2,r}$ and $t_{1/2,d}$ have been simply 
discussed to state that one transient should have similar variability properties as those of the other optical TDEs, although 
the candidates of optical TDEs have very different profiles of observed light curves with very different time durations and very 
different peak intensities, etc..

	Based on the correlation between $t_{1/2,r}$ and $t_{1/2,d}$ (the values listed in Table~4 in \citealt{yr23}) for the 
optical TDEs shown in \citet{yr25} (see the results in the left panel of their Fig.~10), the ratio $R_{1/2,rd}$ of $t_{1/2,r}$ to 
$t_{1/2, d}$ could be a similar value (the slope of the correlation between $t_{1/2,r}$ and $t_{1/2, d}$) for optical TDEs. In 
other words, although optical TDEs have very different values of $t_{1/2,r}$ and $t_{1/2,d}$ (from several days to more than a 
hundred days), they have similar physical processes but with intrinsic stars with different stellar parameters tidally disrupted 
by central SMBHs with different BH masses. Therefore, to check whether the parameter $R_{1/2,rd}$ is a constant value for optical 
TDEs is the main objective of the manuscript.

\begin{figure*}
\centering\includegraphics[width = 18cm,height=4.8cm]{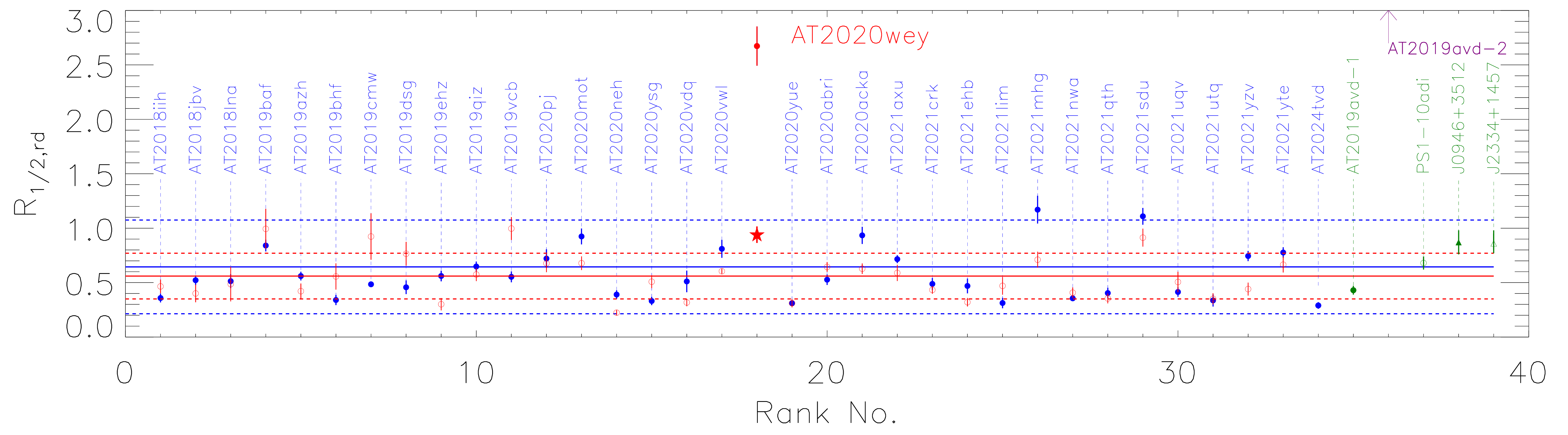}
\caption{Properties of $R_{1/2,rd}$. X-labels from 1 to 34 mark the Rank No. of the 33 optical TDEs in \citet{yr23} and AT2024tvd 
in \citet{yr25}. Solid red circle plus error bars mark the results for AT2020wey. Solid red five-point-star plus error bars mark 
the modified results in AT2020wey. Solid circle in dark green (Rank No.=35) shows the results for the first flare in AT2019avd, 
and the up arrow in purple (Rank No.=36) shows the $R_{1/2,rd}\sim11$ higher than the current marked position for the second flare 
in AT2019avd. Open circle (Rank No.=37), solid triangle (Rank No.=38) and open triangle (Rank No.=39) plus error bars in dark 
green show the results in PS1-10adi, SDSS J0946+3512 and J2334+1457. Open circles plus error bars in red show the re-measured 
$R_{1/2,rd}$ of the 32 TDEs (except AT2020wey) in \citet{yr23}. Horizontal solid and dashed line in blue and in red 
mark the mean value and $\pm$ standard deviation of the $R_{1/2,rd}$ in \citet{yr23} and the re-measured $R_{1/2,rd}$ through 
ZFPS r-band background-subtracted light curves in this manuscript.}
\label{rd}
\end{figure*}

\begin{figure*}
\centering\includegraphics[width = 18cm,height=3.5cm]{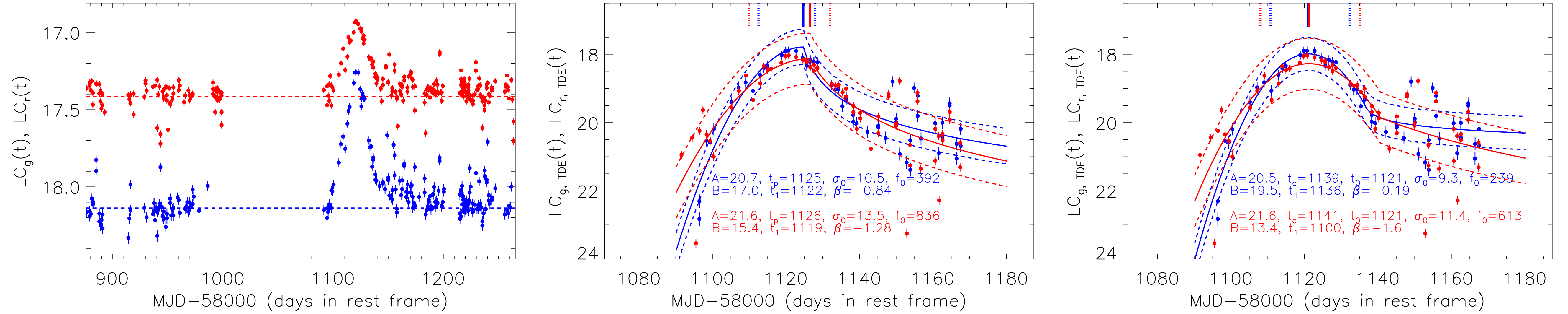}
\caption{Results for AT2020wey. Left panel shows the ZTF gr-band light curves (symbols in blue and in red) of AT2020wey in rest 
frame, with horizontal dashed lines marking the determined apparent magnitudes of the host galaxy contributions. Middle panel 
shows the $LC_{g,TDE}(t)$ (blue symbols) and $LC_{r,TDE}(t)$ (red symbols) after subtractions of the host galaxy contributions. 
In middle panel, solid and dashed lines in blue and in red show the mathematical method determined best descriptions and 
corresponding 1RMS scatters to the $LC_{g,TDE}(t)$ and $LC_{r,TDE}(t)$, respectively. The corresponding parameters are marked in 
blue and red characters. Vertical solid lines in blue and in red mark the peak positions of the best fitting results to the 
$LC_{g,TDE}(t)$ and $LC_{r,TDE}(t)$, the vertical dotted lines in blue and in red mark the corresponding positions for the 
half-max of the best fitting results. Right panel shows the new descriptions to the $LC_{g,TDE}(t)$ and $LC_{r,TDE}(t)$ by the 
modified formula, with symbols and line styles having the same meanings as those in the middle panel.}
\label{lmc}
\end{figure*}

	Based on the reported timescales of $t_{1/2,r}$ and $t_{1/2,d}$ of the 33 optical TDEs in \citet{yr23} and AT2024tvd 
in \citet{yr25}, the ratios $R_{1/2,rd}=t_{1/2,r}/t_{1/2,d}$ are shown in Fig.~\ref{rd}. Except AT2020wey which has firstly 
been discovered by \citet{cp23} and reported with $R_{1/2,rd}\sim2.7$ in \citet{yr23}, the other 33 optical TDEs have similar 
$R_{1/2,rd}\sim0.6$ (standard deviation about 0.4). It is interesting to check whether the larger $R_{1/2,rd}$ in AT2020wey 
is intrinsically different from the other optical TDEs. Therefore, AT2020wey ($z\sim0.0274$) is selected as the target of the 
manuscript. Section 2 presents our main results on timescales of AT2020wey through its high quality light curves from Zwicky 
Transient Facility (ZTF) \citep{bk19, ds20}, and necessary discussions on the other five optical flares. Section 3 gives our 
conclusions. And, we have adopted the cosmological parameters of $H_{0}=70{\rm km\cdot s}^{-1}{\rm Mpc}^{-1}$, 
$\Omega_{\Lambda}=0.7$ and $\Omega_{\rm m}=0.3$.

\section{Main Results and necessary discussions}

	The ZTF gr-band light curves $LC_g(t)$, $LC_r(t)$ in magnitude space of \obj are shown in left panel of Fig.~\ref{lmc}, 
with MJD-58000 in rest frame from 870 to 1260. The magnitudes are based on the PSF photometry \citep{bk19}. Here, 
ZTF i-band light curve or light curves from UVOT and ATLAS are not considered, due to few data points. Based on the light curves, 
apparent magnitudes $m_H$ of the host galaxy contributions in the gr-band can be determined as 18.13 and 17.41 mags, after 
considering the mean magnitudes of the data points with MJD-58000 smaller than 1000days. After subtractions of the host galaxy 
contributions, the intrinsic variability $LC_{g,TDE}(t)$ and $LC_{r,TDE}(t)$ related to central TDE can be determined by 
$LC_{TDE}(t) = -2.5\log(10^{-0.4 LC(t)} - 10^{-0.4 m_H})$, shown in middle panel of Fig.~\ref{lmc}. Meanwhile, the 
background-subtracted light curves in flux space from the difference imaging of \obj~ are selected from ZTF Forced Photometry 
Service (ZFPS) and discussed in Appendix A.

\begin{figure*}
\centering\includegraphics[width = 18cm,height=3.5cm]{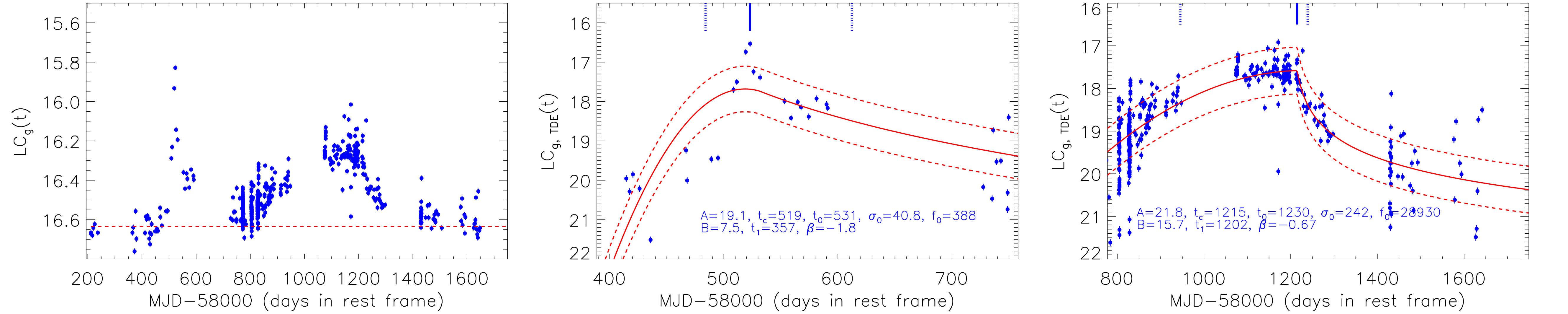}
\caption{Results for AT2019avd. Left panel shows the ZTF g-band light curve in rest frame, with horizontal dashed red line marking 
the determined apparent magnitude of the host galaxy contributions. Middle panel shows the $LC_{g,TDE}(t)$ of the first flare, after 
subtractions of the host galaxy contributions. In middle panel, solid and dashed lines in red show the mathematical method 
determined best descriptions and corresponding 1RMS scatters. The corresponding parameters are marked in blue characters. Vertical 
solid and dotted lines in blue mark the peak and half-maximum positions of the best fitting results. Right panel shows the results 
for the second flare, with symbols and line styles having the same meanings as those in the middle panel.}
\label{cd22}
\end{figure*}

	Then, we firstly check whether the larger $R_{1/2,rd}$ in \obj~ is from unintentionally wrong measurements. Based on the 
Gaussian rise and power-law decay in luminosity space in \citet{yr23}, $LC_{g,TDE}(t)$ and $LC_{r,TDE}(t)$ in magnitude space 
can be described by
\begin{equation}
LC_{TDE}(t)~=~\left\{
\begin{aligned}
	&A~-~2.5\log(G([t_p, \sigma_0, f_0])) \ \ \ (t<t_p) \\
	&B~-~2.5\log((t - t_1)^\beta) \ \ \ \ \ \ \ \ (t>t_p)
\end{aligned}
\right.
\end{equation}
with $t_p$ as time of peak bright and $G([t_p,\sigma_0,f_0])$ as a Gaussian function with $t_p$, $\sigma_0$ and $f_0$ as mean, 
second moment and total area of the desired Gaussian curve. Here, one point should be noted. In the manuscript, without considering 
the additional exponential decline with a secondary peak applied in AT2020wey in \citet{yr23} to describe tails of $LC_{TDE}(t)$, 
only the Gaussian rise and power-law decay are mainly considered and applied to describe the $LC_{TDE}(t)$. The best fitting 
results and the determined model parameters to $LC_{g,TDE}(t)$ and $LC_{r,TDE}(t)$ are shown in the middle panel of Fig.~\ref{lmc}, 
through the Levenberg-Marquardt least-squares minimization technique \citep{mc09}. The $t_{1/2,r}$ and $t_{1/2,d}$ are determined 
as 12.2$\pm$0.5 days and 3.2$\pm$0.4 days, 16.7$\pm$0.9 days and 5.5$\pm$0.5 days in gr-band. The uncertainties are determined 
through the 1RMS scatters of the best fitting results. In magnitude space, the half-maximum positions are determined by positions 
to be 0.7526 mags ($2.5\log(2)$) plus the minimum values of the best fitting results. The re-measured timescales are similar as 
those in \citet{yr23} in AT2020wey, indicating larger $R_{1/2,rd}$ in AT2020wey not from wrong measurements.

	The very different $R_{1/2,rd}$ in AT2020wey from the other 33 optical TDEs probably indicates different intrinsic physical 
process in AT2020wey from the processes for the other optical TDEs. Therefore, it is necessary to check whether are there effects 
for modifying $R_{1/2,rd}$ in AT2020wey to be similar as the values in the other 33 optical TDEs. If yes, the parameter $R_{1/2,rd}$ 
could be a potential classification parameter for optical TDEs.

	Based on the functions above, the determined peak positions have strong effects on the determined timescales. Therefore, 
slightly modified but more flexible formula can be given as 
\begin{equation}
LC_{TDE}(t)~=~\left\{
	\begin{aligned}
		&A~-~2.5\log(G([t_0, \sigma_0, f_0])) \ \ \ (t<t_c) \\
		&B~-~2.5\log((t - t_1)^\beta) \ \ \ \ \ \ \ \ (t>t_c)
	\end{aligned}
	\right.
\end{equation}
with $t_c$ not fixed to $t_p$ but a free parameter and $G([t_0, \sigma_0, f_0])$ meaning a Gaussian function with $t_0$ not fixed 
to $t_p$ or $t_c$). The new formula can lead to new descriptions to $LC_{g,TDE}(t)$ and $LC_{r,TDE}(t)$, as shown in right panel 
of Fig.~\ref{lmc}. And the re-determined $t_{1/2,r}$ and $t_{1/2,d}$ are 10.3$\pm$0.6 days and 11.2$\pm$0.5 days, 13.1$\pm$0.5 days 
and 13.9$\pm$0.2 days in gr-band. The re-determined $t_{1/2,r}$ and $t_{1/2,d}$ are 11.7$\pm$0.6 days and 12.6$\pm$0.4 days (the 
mean values in gr-band) in AT2020wey. Now, the results of AT2020wey with $R_{1/2,rd}\sim0.94\pm0.08$ are re-plotted in Fig.~\ref{rd}, 
leading AT2020wey to have $R_{1/2,rd}$ similar as the other optical TDEs.

	Moreover, the flexible model functions have also been applied to describe the gr-band background-subtracted light curves 
in flux space of the other 32 TDEs in \citet{yr23} in Appendix B, leading to the re-measured $R_{1/2,rd}$ in r-band shown as open 
red circles in Fig.~\ref{rd}. As discussed in Appendix B, through g-band and r-band light curves, there are the same final 
conclusions on properties of $R_{1/2,rd}$. There are no considerations of physical process to describe the optical light curves, 
but the mathematical method determined $R_{1/2,rd}$ have the mean values about 0.60 (standard deviation 0.4) and 0.56 (standard 
deviation 0.21) based on the $R_{1/2,rd}$ in g-band in \citet{yr23} and on our re-measured $R_{1/2,rd}$ in r-band (same results 
in g-band). The smaller standard deviation indicates that applications of the flexible model functions are robust and preferred. 
And, the results above can also be applied to confirm that to measure values or mean values of $R_{1/2,rd}$ through ZTF or ZFPS 
single-band or multiple-band light curves in flux space or in magnitude space has few effects on our final results. The similar 
$R_{1/2,rd}$ indicates the optical TDEs from ZTF have probably similar intrinsic physical process for their optical light curves. 
In other words, optical transients if have very different physical processes from the process for optical TDEs, very different 
$R_{1/2,rd}$ could be expected.

\begin{figure*}
\centering\includegraphics[width = 18cm,height=3.5cm]{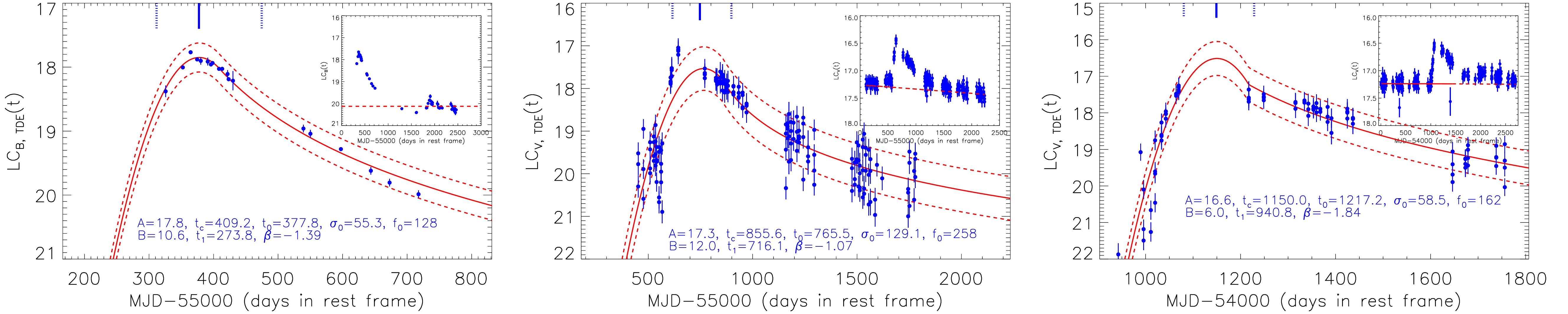}
\caption{Results for the transients of PS1-10adi (left), SDSS J0946+3512 (middle) and J2334+1457 (right). Top right corner of each 
panel shows the light curve in rest frame, with horizontal dashed red line marking the determined apparent magnitude of host galaxy 
contributions. In main body of each panel, solid and dashed lines in red show the best descriptions and corresponding 1RMS scatters 
to the $LC_{TDE}(t)$, after subtractions of the host galaxy contributions, with symbols and line styles have the same meanings as 
those in the right panel of Fig.~\ref{cd22}.}
\label{kk17}
\end{figure*}

	Furthermore, we check whether the other optical transients have similar $R_{1/2,rd}$ as the 34 optical TDEs discussed above. 
Here, the following optical transients are mainly considered. \citet{cd22} have discussed the different physical origins of the two 
optical flares in AT2019avd ($z\sim0.029$), a stream circularization for the first flare, but the delayed BH accreting process for 
the second flare. Left panel of Fig.~\ref{cd22} shows the ZTF g-band light curve of AT2019avd with apparent magnitude 16.63 mags 
from the host galaxy contributions. Here, because number of data points in the ZTF r-band light is only half of that in ZTF g-band 
light curve, the ZTF r-band light curve is not considered in AT2019avd. Then, the flexible model functions in Equation (2) are 
applied to describe the light curves of the two flares after subtractions of the host galaxy contributions, leading to the best 
descriptions shown in the middle and right panels of Fig.~\ref{cd22}. The determined $t_{1/2,r}$ and $t_{1/2,d}$ are 38.9$\pm$1.0 
days and 89.4$\pm$4.9 days for the first flare, and 240.0$\pm$2.9 days and 23.3$\pm$1.9 days for the second flare. Therefore, the 
first flare and the second flare have $R_{1/2,rd}$ to be 0.43$\pm$0.04 and 11.4$\pm$1.2, also shown in Fig.~\ref{rd}. The first 
flare has similar intrinsic physical process as those in the other optical TDEs, but the second flare probably has very different 
physical process from the other optical TDEs. To discuss the different physical origins of the two optical flares in AT2019avd is 
beyond the scope of the manuscript, therefore, there are not further discussions any more on physical models for the two optical 
flares in AT2019avd.

	Meanwhile, PS1-10adi ($z\sim0.203$) is one another target to be discussed, because \citet{kk17} have reported PS1-10adi 
as a highly energetic transient event probably different from common optical TDEs. The flexible model functions in Equation (2) 
are applied to describe the B-band light curve $LC_{B,TDE}(t)$, after subtractions of the host galaxy contributions with apparent 
magnitude 20.12 mags. The best fitting results and the 1RMS scatters are shown in the left panel of Fig.~\ref{kk17}, leading to 
the determined $t_{1/2,r}=65.3\pm1.7$ days and $t_{1/2,d}=96.8\pm5.8$ days. Therefore, PS1-10adi has $R_{1/2,rd}=0.68\pm0.06$, 
similar as the other common optical TDEs (except the second flare in AT2019avd), indicating similar intrinsic physical process 
in PS1-10adi as those in other optical TDEs. Then, the other two reported PS1-10adi-like transients of SDSS J0946+3512 ($z\sim0.119$) 
and J2334+1457 ($z\sim0.107$) in \citet{kk17} are also discussed and shown in the middle and right panels of Fig.~\ref{kk17} 
with their V-band light curves selected from Catalina Sky Survey \citep{dr09}. The corresponding $R_{1/2,rd}$ are 0.87$\pm$0.11 
($t_{1/2,r}$ and $t_{1/2,d}$ to be 131.4$\pm$7.1 days and 151.9$\pm$8.9 days) and 0.86$\pm$0.09 ($t_{1/2,r}$ and $t_{1/2,d}$ to 
be 68.7$\pm$2.7 days and 879.5$\pm$5.4 days) in SDSS J0946+3512 and J2334+1457, respectively, similar as the other optical TDEs 
(except the second flare in AT2019avd) as shown in Fig.~\ref{rd}. Here, the reported transients of PS1-13jw, CSS100217 and SDSS 
J0948+0318 in \citet{kk17} are not considered, because that we can not find good light curve of PS1-13jw, and that there are 
apparent effects of obscurations leading to observed variability profile very different from the intrinsic variability profile 
in CSS100217 as our recently discussed in \citet{gz25}, and that there are apparent variability components (probably due to 
intrinsic AGN variability) around the flare in SDSS J0948+0318.

	Before ending the section, properties of $R_{1/2,rd}$ can be simply checked in supernova-like (SN-like) transients. Based 
on recent discussions in \citet{ds24} (see their Fig.~4), as discussed in Appendix C, SN-like transients have basically different 
properties of $R_{1/2,rd}$ from those in TDEs. Meanwhile, through the correlation between rise time (half peak to peak) 
and fade time (peak to half peak) in \citet{pf20} (see their Fig.~3) for a large sample of SN-like transients, ratios of rise time 
to fade time have statistical mean values larger than 1, but with the measurements of rise and fade time including host galaxy 
contributions. Based on the discussions in Appendix C and the results in \citet{pf20}, some individual SN-like transients might 
also have the same ratios of $R_{1/2,rd}$, but SN-like tansients should have different statistical properties of $R_{1/2,rd}$ from 
those in TDEs. In the near future, a manuscript is being prepared to provide more detailed results of $R_{1/2,rd}$ for more 
optical TDEs and the other kinds of optical transients, to further support whether $R_{1/2,rd}$ is an efficient parameter to test 
similar and/or different physical processes in different optical TDEs or in different types of transients. However, based on the 
discussed $R_{1/2,rd}$ at current stage for the 39 optical flares, except the second flare in AT2019avd, the other flares have 
the similar $R_{1/2,rd}$, indicating the parameter $R_{1/2,rd}$ should be a potential efficient parameter to classify optical 
flares only through observational variability profiles.

\section{Conclusions}

	Motivated by the timescale correlation between $t_{1/2,r}$ and $t_{1/2,d}$, the parameter $R_{1/2,rd}$ is proposed 
to classify optical TDEs only through variability profiles of optical light curves, especially based on the 34 optical TDEs 
from ZTF having similar $R_{1/2,rd}$ around 0.6. And the $R_{1/2,rd}$ have been checked in the reported four unique transient 
events. We can find that PS1-10adi, SDSS J0946+3512 and J2334+1457 and the first flare in AT2019avd have their $R_{1/2,rd}$ 
similar as the ones of the other optical TDEs, but the second flare in AT2019avd has its $R_{1/2,rd}$ vary larger than the 
values for the other optical TDEs. The results can be applied to potentially confirm that the two flares in AT2019avd have 
intrinsically different physical process, but the PS1-10adi, SDSS J0946+3512 and J2334+1457 and the first optical flare in 
AT2019avd have similar physical processes as those of the other optical TDEs.

\begin{acknowledgements}
Zhang gratefully acknowledge the anonymous referee for giving us constructive comments to greatly improve the 
paper. Zhang gratefully thanks the kind grant support from Guangxi Science and Technology Program 2026GXNSFDA00640018, and 
from NSFC-12373014, 12173020 and the support from Guangxi Talent Programme (Highland of Innovation Talents). 
This manuscript has made use of the data from ZTF, CSS and MPFIT package. 
\end{acknowledgements}


\appendix

\section{light curves of \obj~ from the difference imaging}

\begin{figure}
\centering\includegraphics[width = 8cm,height=9.5cm]{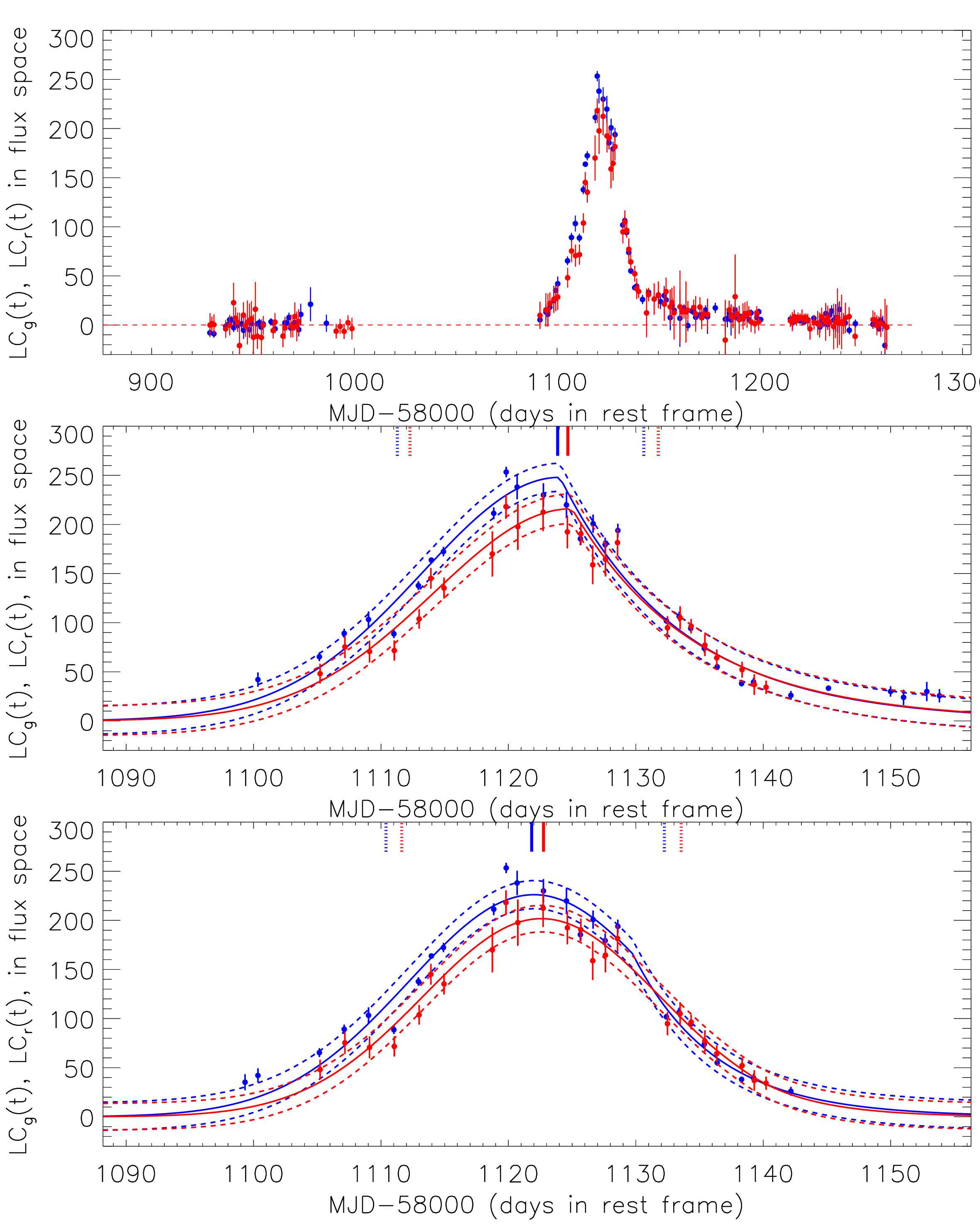}
\caption{Results on background-subtracted ZFPS light curves of AT2020wey in rest frame. Top panel shows the ZFPS gr-band
background-subtracted light curves (symbols in blue and in red) in flux space, including the data points with measurements smaller 
than 3 times of their uncertainties. Horizontal dashed red line marks flux equal to zero. Middle panel and bottom panel show the 
best fitting results to the background-subtracted light curves with measurements larger than 3 times of their uncertainties around 
the TDE related flare with MJD-58000 between 1100 and 1200, with symbols and line styles having the same meanings as those in the 
middle panel and right panel in Fig.~\ref{lmc}.}
\label{zfps}
\end{figure}

	Besides the shown light curves in Fig.~\ref{lmc} selected from ZTF, the background-subtracted light curves from the 
difference imaging of \obj~ are also selected from ZTF Forced Photometry Service (ZFPS, 
\url{https://ztfweb.ipac.caltech.edu/batchfp.html}, registered username and password required) and shown in Fig.~\ref{zfps} in 
flux space. Then, similar as done in middle panel and right panel of Fig.~\ref{lmc}, based on the corresponding equations in 
flux (luminosity) space in \citet{yr23} and the flexible equations in flux space 
\begin{equation}
LC_{TDE}(t)~=~\left\{
        \begin{aligned}
                &G([t_0, \sigma_0, f_0])) \ \ \ (t<t_c) \\
		&B~\times~exp[(t - t_1)/\tau] \ \ \ (t>t_c)
        \end{aligned}
        \right.
\end{equation}
with $t_c$ not fixed to $t_p$ but a free parameter, the best fitting results and corresponding 1RMS scatters can be determined 
to the background-subtracted ZFPS light curves, and shown in middle panel and bottom panel of Fig.~\ref{zfps}. Based on the 
equations in \citet{yr23}, the determined $t_{1/2, r}$ and $t_{1/2,d}$ are 12.6$\pm$1.1 days, 6.8$\pm$1.1 days and 12.4$\pm$1.3 
days, 7.1$\pm$1.4 days in gr-band, totally similar as those in \citet{yr23}. And based on the more flexible equations, the 
determined $t_{1/2, r}$ and $t_{1/2,d}$ are 10.6$\pm$0.9 days, 11.1$\pm$1.1 days and 11.2$\pm$0.9 days 10.5$\pm$1.0 days in 
gr-band, leading $R_{1/2,rd}$ to be 1.01$\pm$0.2 similar as the results determined in the magnitude space as shown in right 
panel of Fig.~\ref{lmc}.

	The results above can be applied to confirm that, ZTF light curves and background-subtracted ZFPS light curves can lead 
to totally similar results on determined $R_{1/2,rd}$ in AT2020wey. Meanwhile, for us, light curves from photometric CCD image 
(ZTF light curve) and ZFPS background-subtracted light curves have their own strengths and weaknesses. However, we are more 
accustomed to using the ZTF light curves, due to their clear information of component not related to TDEs. Therefore, in the 
manuscript, rather than the background-subtracted ZFPS light curves discussed in the Appendix, the common ZTF light curves with 
PSF magnitudes determined through the CCD images are selected from 
\url{https://irsa.ipac.caltech.edu/cgi-bin/Gator/nph-scan?submit=Select&projshort=ZTF} (no username or password required) and 
mainly considered in AT 2020wey. More detailed discussions (including technique details) on common ZTF light curves and ZFPS light 
curves can be found in \url{https://irsa.ipac.caltech.edu/Missions/ztf.html}.

\section{re-determined $R_{1/2,rd}$ in the other 32 TDEs by the flexible model functions}

\begin{figure*}
\centering\includegraphics[width = 15.5cm,height=20cm]{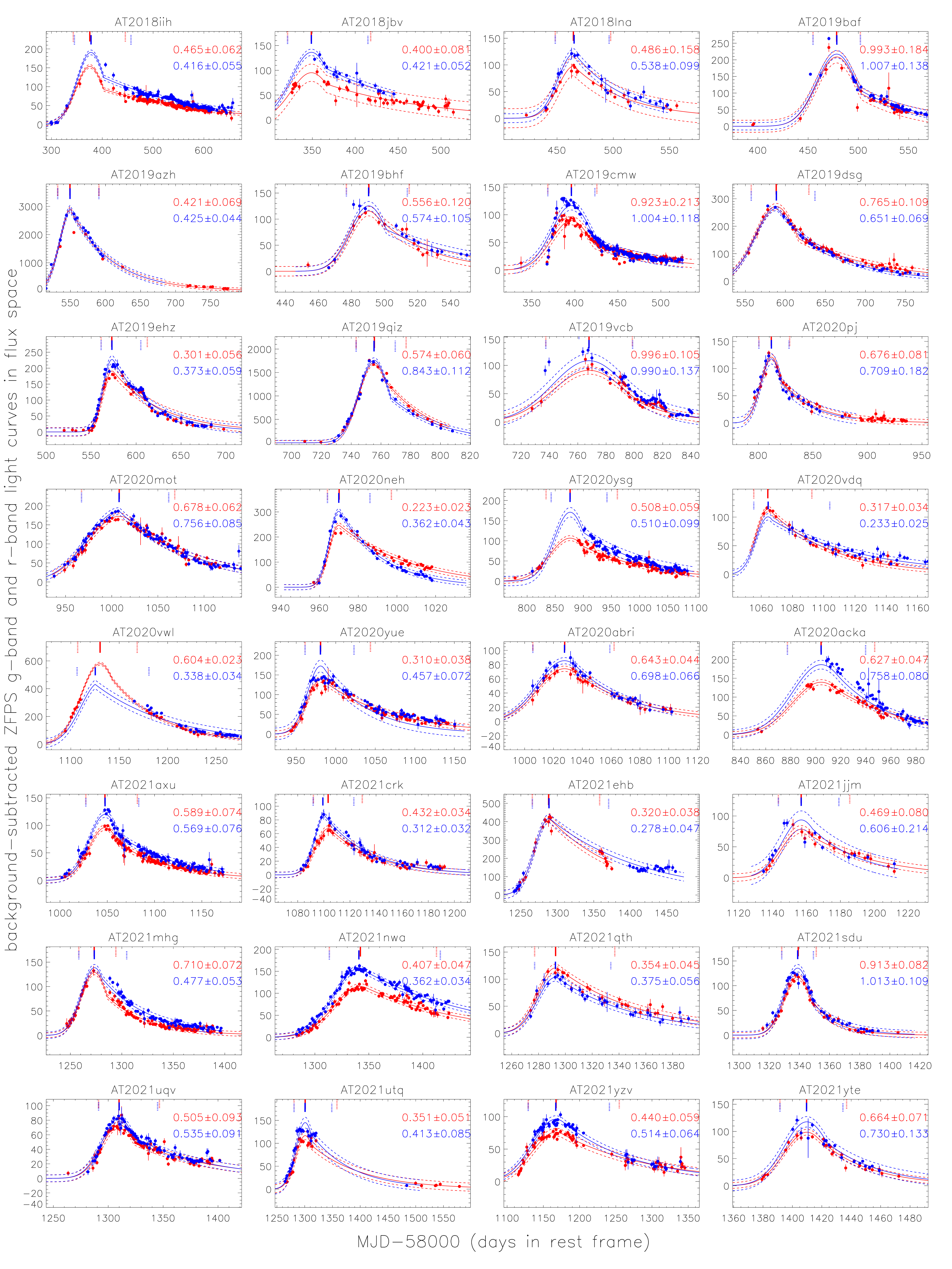}
\caption{Results on background-subtracted ZFPS gr-band light curves of the 32 TDEs in rest frame. In each panel, solid 
circles plus error bars in blue and in red represent the data points in the gr-band. Solid line and dashed lines in blue and in 
red show the best fitting results and corresponding 1 RMS scatters to the g-band and r-band light curves. Vertical solid and 
dashed lines in blue and in red mark the positions of the maximum and half-maximum of the g-band and r-band light curves. The 
determined $R_{1/2,rd}$ are listed in blue and red characters in each panel through the best fitting results to the g-band and 
r-band light curves.}
\label{lmc33}
\end{figure*}

\begin{figure}
\centering\includegraphics[width = 8cm,height=4.5cm]{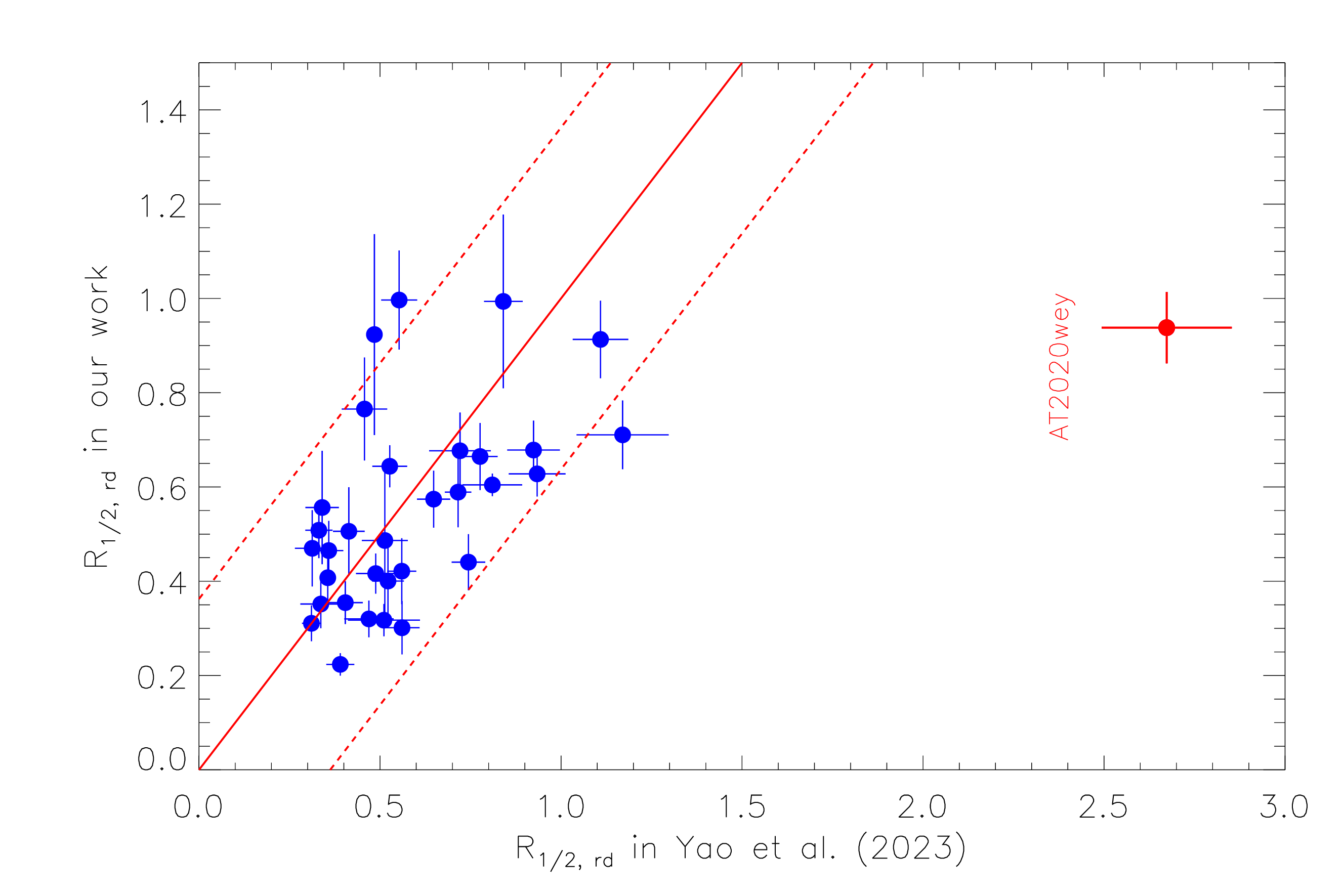}
\centering\includegraphics[width = 8cm,height=4.5cm]{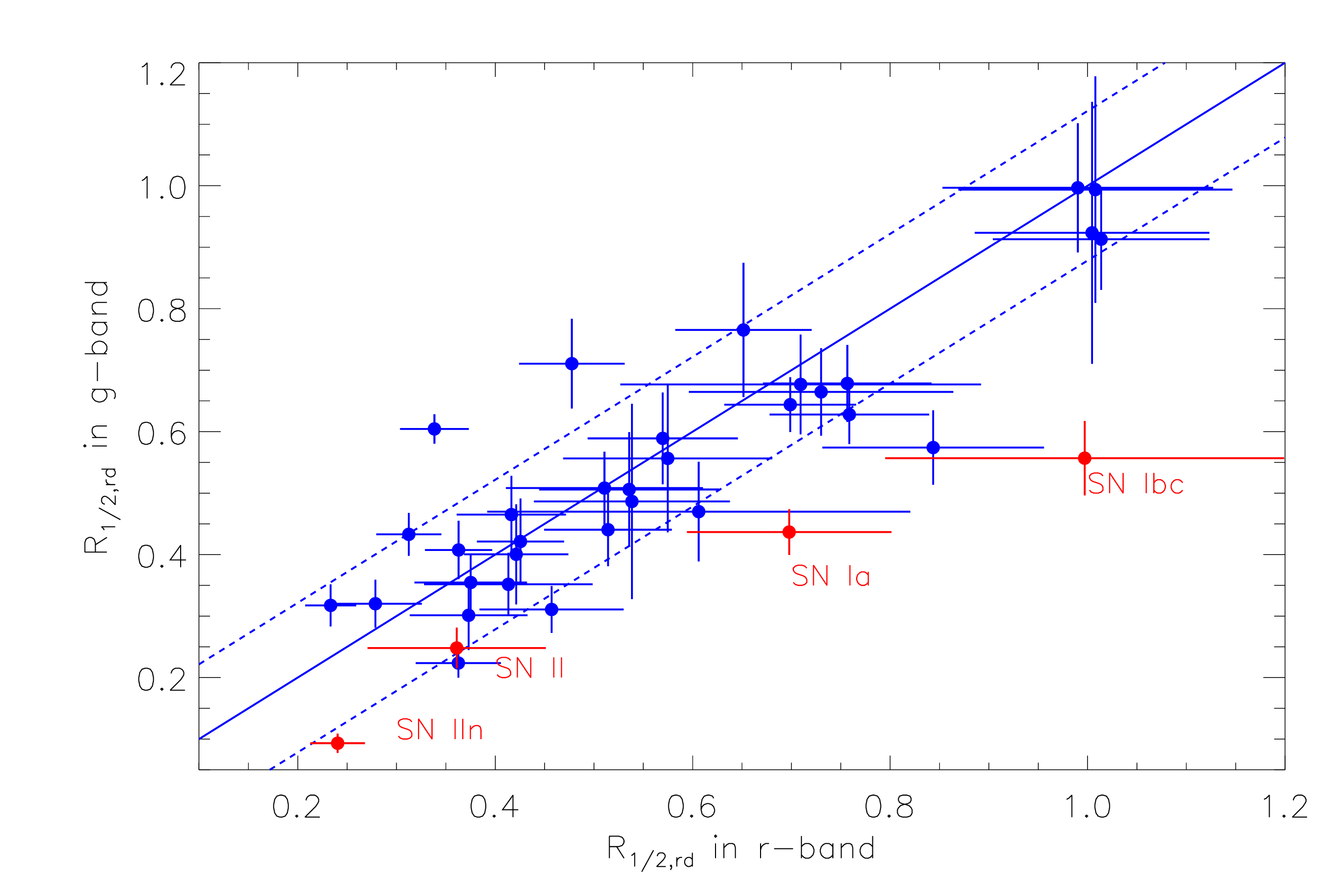}
\caption{Properties of our measured $R_{1/2,rd}$. Top panel shows the correlation between $R_{1/2,rd}$ in \citet{yr23} and our 
re-determined $R_{1/2,rd}$ by the flexible equations applied in r-band in this manuscript for the 33 TDEs. Solid circle plus error 
bars in red show the results for AT2020wey. Solid line and dashed lines in red show $X=Y$ and corresponding 1RMS scatters. Bottom 
panel shows the correlation between $R_{1/2,rd}$ in g-band and $R_{1/2,rd}$ in r-band of the 33 TDEs (solid circle plus error 
bars in blue) and of the 5 classified SN-like transients (solid circles plus error bars in red). Here, due to SLSN-I with 
$R_{1/2,rd}$ around 10, the SLSN-I is not covered in the bottom panel. In bottom panel, solid and dashed blue lines show $X=Y$ 
and corresponding 1RMS scatters for the 33 TDEs.}
\label{lmc34}
\end{figure}

	In order to test robustness of applications of the flexible model functions, the flexible model equations in equation (2) 
have also been applied to describe the background-subtracted ZFPS gr-band light curves in flux space of the other 32 TDEs in 
\citet{yr23}, leading to the re-measured $R_{1/2, rd}$. Fig.~\ref{lmc33} shows the best fitting results and corresponding 1RMS 
scatters to the background-subtracted ZFPS gr-band light curves of the 32 TDEs. Then, top panel of Fig.~\ref{lmc34} shows the 
correlation between the $R_{1/2,rd}$ in \citet{yr23} in g-band and our re-measured $R_{1/2,rd}$ in r-band, with Spearman rank 
correlation coefficient is 0.59 ($P_{null}\sim3\times10^{-4}$). Bottom panel of Fig.~\ref{lmc34} shows the strong linear 
correlation between our re-measured $R_{1/2,rd}$ in r-band and our re-measured $R_{1/2,rd}$ in g-band of the TDEs in \citet{yr23}, 
leading to ratio 0.99 (0.12 as the corresponding 1RMS scatter of X=Y) of re-measured $R_{1/2,rd}$ in r-band to re-measured 
$R_{1/2,rd}$ in g-band. Therefore, through g-band and r-band light curves, there are the same final conclusions on properties 
of $R_{1/2,rd}$.

	Here, without considering the power-law rise, only the Gaussian rise is applied, due to the following two points. First, 
the two rise functions can lead to similar fitting results to the TDEs light curves around half-maximum (see Fig.~8 in \citet{yr23}). 
Second, applications of Gaussian rise can lead to more flexible crossing point of the rise function and the decay function. 
Meanwhile, we did not estimate $A_\nu$ (parameter in Equation 1a and 1b in \citealt{yr23}) by fitting SED of AT2020wey, but 
accepted $A_\nu$ as a free parameter, leading to more flexible fitting results.

	Based on the re-measured $R_{1/2, rd}$, the mean value is 0.56 with 0.21 as the corresponding standard deviation. Compared 
with the mean value 0.6 with 0.4 as the corresponding standard deviation through the values in \citet{yr23}, the smaller standard 
deviation can be applied to support robustness of applications of the flexible model functions.

\section{Properties of $R_{1/2,rd}$ in the five classified SN-like transients}

\begin{figure}
\centering\includegraphics[width = 8.5cm,height=20cm]{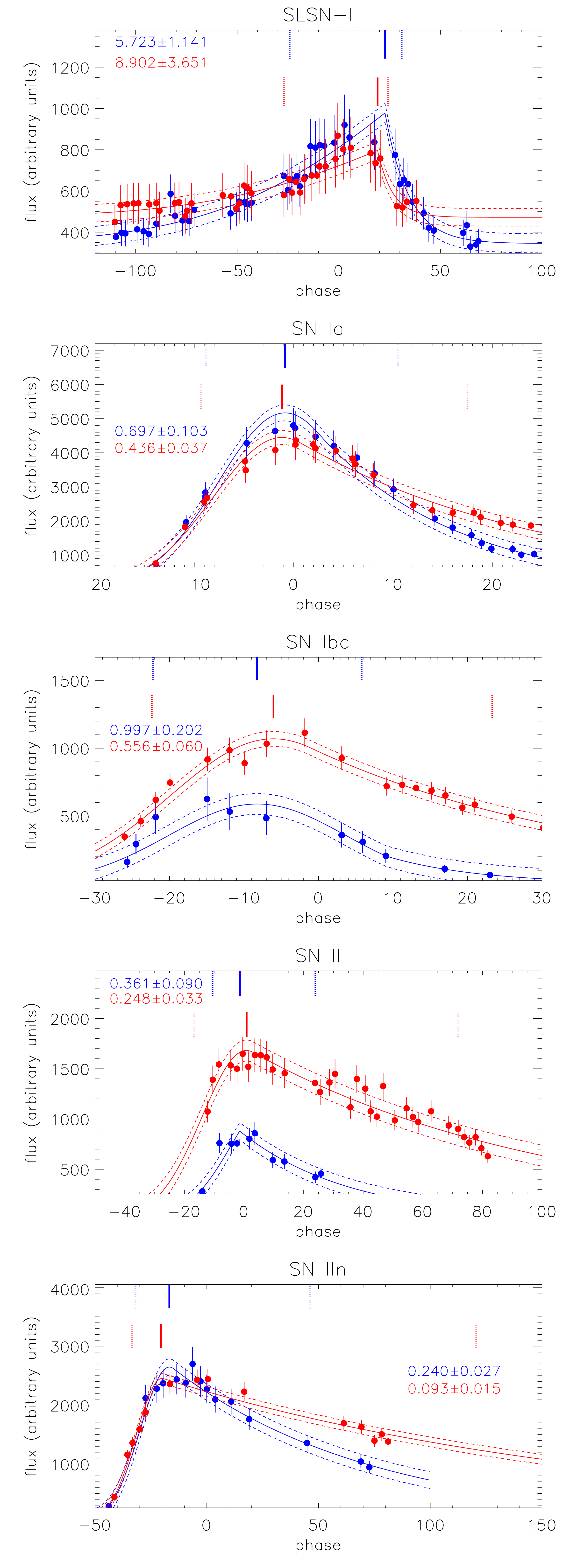}
\caption{Results on the gr-band light curves of the five classified SN-like transients. In each panel, solid circles plus error 
bars in blue and in red represent the data points in g-band and r-band. Solid line and dashed lines in blue and in red show the 
best fitting results and corresponding 1RMS scatters to the g-band and r-band light curves. Vertical solid and dashed lines in 
blue and in red mark the positions of the maximum and half-maximum of the best fitting results to the g-band and to the r-band 
light curves. The determined $R_{1/2,rd}$ are listed in blue and red characters through the best fitting results to the g-band 
and r-band light curves.}
\label{sns}
\end{figure}

	Based on the gr-band light curves in flux space shown in Fig.~4 in \citet{ds24} of the five classified SN-like transients 
(SLSN-I, SN Ia, SN Ibc, SN II and SN IIn), the flexible functions have been applied to determine the $R_{1/2,rd}$ in the SN-like 
transients. Fig.~\ref{sns} shows the best fitting results to the SN-like transients light curves, with the determined $R_{1/2,rd}$ 
listed in each panel. Meanwhile, bottom panel of Fig.~\ref{lmc34} shows the correlation between $R_{1/2,rd}$ in g-band and 
$R_{1/2,rd}$ in r-band of the 5 classified SN-like transients. Then, the following properties of $R_{1/2,rd}$ can be found in the 
SN-like transients.

	In the SNSL-I, two points can be found. First, the determined $R_{1/2,rd}$ in both g-band light curve and r-band light 
curve are apparently larger than 1. Second, there are apparent color evolutions in SNSL-I, leading to $R_{1/2,rd}$ in the g-band 
light curve being different from that in the r-band light curve. The two points can be applied to confirm that the SNSL-I 
transients have very different properties of $R_{1/2,rd}$ from those in standard TDEs.

	In the SN Ia, SN Ibc and SN II, also two points can be found. First, the determined $R_{1/2,rd}$ in both g-band light curve 
and r-band light curve are smaller than 1, and around the mean value of $R_{1/2,rd}$ in standard TDEs. Second, apparent color 
evolutions in SN Ia lead to $R_{1/2,rd}$ in the g-band light curve being different from that in the r-band light curve. Therefore, 
as shown in bottom panel of Fig.~\ref{lmc34}, in the space of $R_{1/2,rd}$ in r-band and $R_{1/2,rd}$ in g-band, SN Ia, SN Ibc and 
SN II are lying in the lower area than the standard TDEs.

	In the SN IIn, the apparent color evolutions can lead $R_{1/2,rd}$ in the g-band light curve to be about 2.6 times of 
$R_{1/2,rd}$ in the r-band light curve. The very different $R_{1/2,rd}$ in light curves in different optical bands can be clearly 
applied to state that there are very different properties of $R_{1/2,rd}$ in SN IIn from those in standard TDEs.

	In one word, based on properties of $R_{1/2,rd}$ through multi-band optical light curves, there are basic clues to support 
different properties of $R_{1/2,rd}$ in SN-like transients from those in standard TDEs. 

\end{document}